\documentclass[aps,prl,preprint,showpacs,superscriptaddress,longbibliography]{revtex4-1}

\usepackage{pgffor}
\usepackage{graphicx}  % needed for figures
\usepackage{dcolumn}   % needed for some tables
\usepackage{bm}        % for math
\usepackage{amssymb}   % for math
\usepackage{amsmath}
\usepackage{esint}
\usepackage{hyperref}
\newcommand{\cl}{black} %replace red by black to eliminate all red, vice versa

\usepackage{lipsum}
\usepackage{titlesec}

\titlespacing\section{0pt}{12pt plus 4pt minus 2pt}{10pt plus 2pt minus 2pt}
\titlespacing\subsection{0pt}{12pt plus 4pt minus 2pt}{5pt plus 2pt minus 2pt}
\titlespacing\subsubsection{0pt}{12pt plus 4pt minus 2pt}{0pt plus 2pt minus 2pt}

\hypersetup{%
   colorlinks=true,
   urlcolor=blue,
   linkcolor=blue,
   citecolor= blue,
   raiselinks=false
}

\begin{document}

\title{Energy Transfer into Period-Tripled States in Coupled Electromechanical Modes at Internal Resonance}% Force line breaks with \\

 \affiliation{Department of Physics, The Hong Kong University of Science and Technology, Hong Kong, China}%Lines break automatically or can be forced with \\
 \affiliation{William Mong Institute of Nano Science and Technology, The Hong Kong University of Science and Technology, Clear Water Bay, Kowloon, Hong Kong, China}

 \affiliation{Department of Physics and Astronomy, Michigan State University, East Lansing, Michigan 48824, USA}

\author{Yingming Yan$^1$$^,$$^2$}
 \author{X. Dong$^1$$^,$$^2$}%
 \author{L. Huang$^1$$^,$$^2$}%
 \author{K. Moskovtsev$^3$}
 \author{H. B. Chan$^1$$^,$$^2$}\thanks{hochan@ust.hk}

\date{\today}% It is always \today, today,

\begin{abstract}

	Efficient energy transfer often occurs between oscillation modes in a resonator when they are tuned to internal resonance.  We design  the eigenfrequencies of two vibrational modes of an electromechanical resonator to be close to a ratio of 3:1 and demonstrate that the energy supplied to the upper mode can be controllably transferred to the lower mode. With the lower mode vibrating with a period tripled that of the upper mode, the discrete time-translation symmetry \color{\cl} imposed by \color{black} the periodic drive is broken. The lower mode settles into one of three stable period-tripled states with different phases. This channel for energy transfer from the upper mode can be turned on or off without changing system parameters. When the upper mode itself becomes multistable under strong resonant or parametric drive, additional sets of coexisting period-tripled states emerge in the lower mode. In the latter case, we measure a total of 6 coexisting vibration states with identical amplitude but phases differing by $\pi$/3. Excitation of coexisting states with three different phases could open new opportunities in designing mechanical memory based on ternary logic. Coupled resonators with period-tripled states can also be used to model complex interacting systems with spin equals one.

\begin{description}
\item[Subject Areas]
nonlinear dynamics, mechanics
\end{description}
 \end{abstract}

                              %display

\maketitle

 \section{I. INTRODUCTION}

 Nonlinear mode coupling enables energy exchange between oscillatory modes with significantly different resonant frequencies. When the resonant frequencies are incommensurate or far from each other, application of parametric modulation is often necessary to enhance the coupling. The enhanced interaction allowed mechanical resonators to be controlled through their coupling with optical, microwave or phononic cavities \cite{kippenberg2008cavity,aspelmeyer2014cavity}. When the frequency of one mode becomes commensurate with an integer multiple of another mode, nonlinear coupling of the modes becomes strong, giving rise to efficient energy transfer and internal resonance, which profoundly affects the response of both modes. Internal resonances in macroscopic systems with mode frequencies in the ratios of 3:1 or 2:1 exhibit a wealth of phenomena including nonlinear oscillations, amplitude-modulated oscillations, multi-stable behavior, dissipative nonlinearities and chaotic motions \cite{alfriend1971stability,ribeiro1999non,blasius1999complex,jiang2005construction,ghayesh2013nonlinear,Kirkendall2016}. Recently, internal resonance has been demonstrated to play an important role in micro- and nano-mechanical resonators \cite{antonio2012frequency,eichler2012strong,samanta2015nonlinear,mangussi2016internal,shaw2016periodic,shoshani2017anomalous,chen2017direct,guttinger2017energy}. When the lower mode is excited by a periodic drive, energy exchange with the upper mode leads to novel behaviors in the steady state response as well as the decay of vibrations when the energy supply is removed. For example, the frequency fluctuations in the self-sustained oscillations of the lower mode of a micromechanical device is dramatically reduced at internal resonance with a higher mode \cite{antonio2012frequency}. In resonators made of nanotube or two-dimensional materials, complex lineshapes and multiple hysteresis are observed \cite{eichler2012strong,samanta2015nonlinear}. The energy exchange between the two modes at internal resonance also leads to anomalous decay of vibrations with rates that changes with energy in both micromechanical structures \cite{chen2017direct} and suspended graphene \cite{guttinger2017energy}.

\begin{figure}
\includegraphics[width=3.4in]{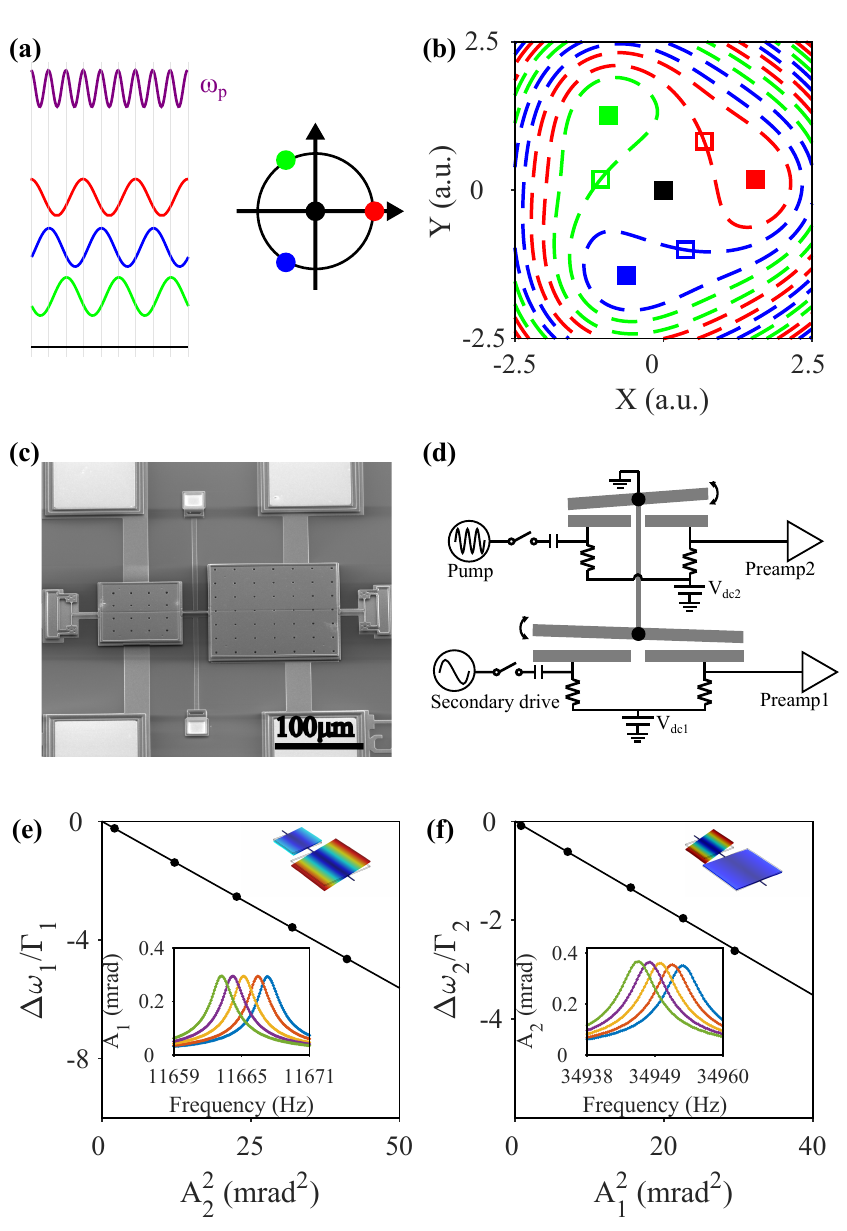}
\caption{(a) Periodic modulation of the cubic term in the confinement potential of a single mode at frequency ${\omega}_{f}$ close to 3${\omega}_{o}$ generates four coexisting stable states. One of them (black) is a zero-amplitude state. The other three (red, blue and green) have identical, non-zero amplitude, with period tripled that of the modulation. They are phase shifted by 2$\pi$/3 relative to one another. (b) The phase space is three-fold symmetric, with four stable states shown as solid squares and the saddle points shown as hollow squares. The separatrices, represented by dashed lines, go through the latter. (c) Scanning electron micrograph of the device consisting of two suspended plates connected by a torsional rod. (d) Excitation and detection scheme (not to scale). (e) The dependence of the scaled shifts of the frequency of mode 1 on the square of the vibration amplitude A${}_{2}$ of mode 2. The line is a linear fit. Lower inset: the spectra of mode 1 in response to a small periodic probe voltage for the data points in the main figure. Upper inset: the vibration profile of mode 1 involves mainly the rotations of the large plate. (f) A similar plot for mode 2 where mostly the small plate rotates.}
\label{fig01} 
\end{figure}

 While much progress has been made by pumping energy into the lower mode, the regime in which energy is supplied to the upper mode for systems at internal resonance is much less explored. When the nonlinear coupling opens up a channel of energy transfer from the upper mode to the lower mode, stable vibrations are induced in the latter with a period that is an integer multiple of the period of the drive. As a result, the discrete time-translation symmetry \cite{dykman2018interaction,eichler2018parametric} \color{\cl} imposed by \color{black} the periodic drive is broken. The breaking of discrete time translation symmetry is well-known in single mode parametric oscillators \cite{rugar1991mechanical,mahboob2008bit,karabalin2010efficient,mahboob2011interconnect,villanueva2011nanoscale,eichler2011parametric,mahboob2014multimode,poot2014classical,seitner2017parametric,goto2018boltzmann,nosan2019gate}, where vibrations are period-doubled relative to the modulation. The two coexisting stable oscillation states with identical amplitude but opposite phase can be used to represent a classical bit \cite{woo1971fluctuations} or an Ising spin \cite{wang2013coherent,mcmahon2016fully,goto2016bifurcation,inagaki2016coherent,inagaki2016large} for logic or computation applications. For parametrically-driven quantum oscillators, the breaking of time-translation symmetry \color{\cl} imposed by \color{black} the periodic drive, together with many-body interactions and disorder, can lead to the formation of time crystals \cite{khemani2016phase,else2016floquet,yao2017discrete}. While period-doubled oscillations in a single mode rely on modulations of the parabolic term in the confinement potential, it has been predicted \cite{PhysRevA.96.052124,PhysRevResearch.1.023023,PhysRevB.101.054501} that if the potential contains a cubic term that is modulated at a frequency ${\omega}_{f}$ close to 3 times the resonant frequency ${\omega}_{o}$, period-tripled oscillations can be excited to break the discrete time-translation symmetry \color{\cl} imposed by \color{black} the modulation. The simplest model for a single-mode system with stable period-tripled oscillations is given by the equation of motion:
\begin{equation} \label{GrindEQ__1_} 
\ddot{q}+{\omega }^2_0q+2\mathrm{\Gamma }\dot{q}+\left({\gamma }/{m}\right)q^3=({F_{\mathrm{p}}}/{m})q^2\mathrm{cos}\mathrm{}({\omega }_ft) 
\end{equation} 
where \textit{q} is the displacement of the mode, \textit{m} is the effective mass, $\gamma$ denotes the Duffing  nonlinearity and \textit{F${}_{p}$} is the strength of the parametric pump at frequency ${\omega}_{f}$. The system settles into one of three stable vibrations states with phases differing by 2$\pi$/3 or the zero-amplitude state [Figs.~\ref{fig01}(a) and (b), where $q(t)= Xcos(\frac{{\omega }_f}{3}t)+ Ysin(\frac{{\omega }_f}{3}t)$]. Unlike period-doubled oscillations where the coefficient of the $q$ term in the equation of motion is modulated in time, for period-tripled oscillations the time modulation is associated with the $q^2$ term.

 Apart from direct parametric modulation of the mode through a term of $q^2\mathrm{cos}\mathrm{}({3\omega }_1t)$, period-tripled oscillations can also be excited by supplying energy to a higher mode that is at 3:1 internal resonance with the first mode with resonant frequency $\omega_1$. Through the interaction energy ${{\gamma }_{12}q}^3_1q_2$ ($q_{1,2}$ represent the displacements of modes 1 and 2 respectively), oscillations in mode 2 at amplitude A${}_{2}$ at its eigenfrequency ${\omega}_{2}$ generates an effective parametric modulation of the form $3{{\gamma }_{12}q}^2_1\mathrm{[}{{\mathrm{A}}_2\mathrm{cos} \left(3{\omega }_1t\right)\ }]$ in the equation of motion of mode 1. Internal resonance facilitates the transfer of energy from mode 2 to mode 1, opening the possibility to excite oscillations in the latter with a period that is three times the periodic drive. Whether stable period-tripled states can be excited in mode 1 depends on the dynamics of both modes because oscillations in mode 1 also affect mode 2. Enhancement of the drive by a second mode to generate period-tripled oscillations was demonstrated in superconducting resonators \cite{PhysRevB.96.174503}. The signal includes contributions from the period-tripled states and the zero-amplitude state as the system switches among them due to noise. Maintaining the system in a period-tripled state with practically no interstate switching may be beneficial for a number of applications such as memory and frequency conversion. To our knowledge, such stable period-tripled oscillations have not been demonstrated. 

 Here, we show that a properly designed electromechanical resonator with two vibrational modes at frequencies close to 3:1 internal resonance exhibits stable period-tripled vibrations in the lower mode (mode 1) that breaks the discrete time-translation symmetry of a periodic drive applied to the higher mode (mode 2) near its eigenfrequency. Since the zero-amplitude state of mode 1 remains stable for all drive frequencies and amplitudes, energy transfer from mode 2 to create the period-tripled states requires activation. Mode 1 needs to be perturbed by a secondary drive, which is subsequently removed.  By choosing the phase of the perturbation, mode 1 can be controlled to settle into one of the three period-tripled states that differ in phase by 2$\pi$/3 when the periodic drive on mode 2 is above a certain threshold value, but remains relatively weak. At certain system parameters, a set of 3 coexisting limit cycles emerges in the rotating frame. When the drive amplitude on mode 2 is increased beyond another threshold, a second set of period-tripled vibrations at a different amplitude can be excited for mode 1, yielding a total of seven coexisting states for a range of drive frequencies. If the resonant drive on mode 2 is replaced by parametric modulation of the confinement potential near twice its eigenfrequency, we observe six coexisting vibrational states in mode 1 with the same amplitude but with phases differing by integer multiples of $\pi$/3. These states are period-sextupled relative to the parametric modulation on mode 2. Apart from fundamental interests relevant to the breaking of discrete time symmetry and classical time crystals \cite{PhysRevLett.123.124301}, the ability to prepare the system in a specific period-tripled state opens new opportunities in using mechanical resonators in ternary memory and logic \cite{2019Tunnelling}, as well as in simulations of spin systems with spin of 1 \cite{DILORENZO20151681} in a context similar to Ising machines for spin of 1/2 \cite{wang2013coherent,mcmahon2016fully,goto2016bifurcation,inagaki2016coherent,inagaki2016large}.
 
 We emphasize that the stable period-three vibrations described in this paper are qualitatively different from the multiple-period oscillations \color{\cl} often observed in \color{black} strongly nonlinear systems \color{\cl} where the nonlinear part of the energy is of the same order of magnitude as the harmonic part. \color{black} The latter are best known in the context of the period-multiplication cascades of limit cycles leading to dynamical chaos \cite{PhysRevLett.43.1743,AMABILI2000641,Alneamy_2020}. They are generally profoundly nonsinusoidal, with multiple Fourier components. In contrast, the period-tripling phenomena discussed in this paper rely on much weaker nonlinearity. The magnitude of the nonlinear terms in the potential energy of the modes is a factor of $\sim$$10^{-7}$ smaller than the magnitude of the harmonic terms. Vibrations in both modes are near perfectly sinusoidal. As we show, the very occurrence of the onset of these vibrations is a consequence of the interplay of the internal resonance between the modes and the weak damping. Therefore a comparatively weak drive is sufficient to excite them.
\color{\cl} The results presented in this paper are measured well below the threshold at which chaos occur.
\color{black}

\noindent 
\allowbreak

\section{II. ELECTROMECHANICAL MODES AT INTERNAL RESONANCE}

 Figure \ref{fig01}(c) shows a typical device, consisting of two movable plates of different sizes, with dimensions of 160 $\mu$m by 156 $\mu$m by 3.5 $\mu$m and 104 $\mu$m by 100 $\mu$m by 3.5 $\mu$m respectively. The two plates are connected together by a suspended beam (30 $\mu$m by 2 $\mu$m by 2 $\mu$m). On the opposite edge for each of the two plates, there is another suspended beam whose other end is anchored to the substrate. Mode 1 involves torsional vibrations of the larger plate with minimal excitation of the small one. For mode 2, the small plate undergoes torsional vibrations. Since the large plate remains almost stationary, the resonant frequency (${\omega}_{2}$/2$\pi$  $\mathrm{\sim}$ 34953.3 Hz) is significantly higher than that of mode 1 (${\omega}_{1}$/2$\pi$ $\mathrm{\sim}$ 11667.5 Hz). The size of the two plates are chosen so that the ratio of their resonant frequencies is close to 3:1. There are two electrodes underneath each plate, as shown in the schematic in Fig.~\ref{fig01}(d). By adjusting the dc voltages {V}${}_{dc1}$ and V${}_{dc2}$ applied to these electrodes, ${\omega}_{1,2}$ can be fine-tuned for the system to go into and out of internal resonance  (see Appendix A). Previous studies of internal resonance in micro- and nano-mechanical systems were often hampered by the difficulty to independently excite and simultaneously detect the vibrations of the two modes. Our design circumvents such problems. For each plate, an ac voltage applied to the electrode on the left in Fig.~\ref{fig01}(d) generates a periodic electrostatic torque to excite the corresponding mode. Vibrations of the mode are detected by measuring the capacitance change between the other electrode and the plate. All measurements are performed at room temperature at pressure of $\mathrm{<}$ 10${}^{-5}$ torr. The damping constants of modes 1 and 2 are ${\Gamma}_{1}/2\pi$ = 0.93 Hz and ${\Gamma}_{2}/2\pi$ = 2.77 Hz respectively.

 We first characterize the system by applying ac voltages to the electrodes to periodically drive the two modes. Both modes have softening nonlinearity, with Duffing constants of $-8.1\times {10}^{-7} \ kg\ m^2\ s^{-2}$  and $-6.1\times {10}^{-8} \ kg\ m^2 s^{-2}$ respectively. There is strong dispersive coupling between the two modes, with coupling energy  $\frac{1}{2}\widetilde{\gamma }{\theta }^2_1{\theta }^2_2$, where ${\theta}_{i}$ is the rotation angle of the plate i for mode i (i = 1, 2). As shown in Figs. 1(e) and 1(f), the resonant frequency of one mode decreases by an amount proportional to the square vibration amplitude of the other mode. The linear fits yield the constant $\widetilde{\gamma }\ $= $-1.02\times {10}^{-7}\ kg\ m^2\ s^{-2}$. While dispersive coupling is not required to generate period-tripled oscillations, we find that it must be included to obtain agreement between our measurement and theory. For the rest of the paper, energy is only supplied to mode 2. Unless otherwise stated, the periodic drive on mode 1 is turned off.
 
Another relevant term in the interaction energy is of the form ${\gamma q}^3_1q_2$, which becomes important for small mismatch ${\epsilon}_{1}$ = ${\omega}_{2}$/3 - ${\omega}_{1}$ when the two modes are near internal resonance. The equations of motions are given by:
\begin{equation} \label{GrindEQ__2_} 
\left\{ \begin{array}{c}
\ddot{{\theta }_1}+2{\mathit{\Gamma}}_1\dot{{\theta }_1}+{\omega }^2_1{\theta }_1+\frac{{\gamma }_1}{I_1}{\theta }^3_1+3\frac{\gamma }{I_1}{\theta }^2_1{\theta }_2+\frac{\widetilde{\gamma }}{I_1}{\theta }_1{\theta }^2_2=0 \\ 
\ddot{{\theta }_2}+2{\mathit{\Gamma}}_2\dot{{\theta }_2}+{\omega }^2_2{\theta }_2+\frac{{\gamma }_2}{I_2}{\theta }^3_2+\frac{\gamma }{I_2}{\theta }^3_1+\frac{\widetilde{\gamma }}{I_2}{\theta }^2_1{\theta }_2=\frac{{\tau }_d}{I_2}{cos \left({\omega }_dt\right)\ } \end{array}
\right. 
\end{equation} 
where I${}_{1,2}$ are the effective moment of inertia of the two modes and ${\tau}_{d}$ is the amplitude of the periodic torque applied to mode 2. The last and second last terms on the left side of Eqs.~\eqref{GrindEQ__2_} originate from the dispersive coupling and the interaction energy ${\gamma q}^3_1q_2$  respectively. With the driving frequency ${\omega}_{d}$ on mode 2 close to ${\omega}_{2}$, we change from $\theta (t)$ and $\dot{\theta }(t)$ to complex amplitudes ${\theta }_1\left(t\right)=u_1\left(t\right)\mathrm{exp}\ \left[i\left({{\omega }_d}/{3}\right)t\right]+\mathrm{c.c.}$.,  ${\theta }_2\left(t\right)=u_2\left(t\right)\mathrm{exp}\ \left[i{\omega }_dt\right]+\mathrm{c.c.}$. Under the rotating wave approximation \color{\cl} in which the fast-oscillating terms are dropped, \color{black} $u_{1,2}\left(t\right)$ evolves according to:
\begin{equation} \label{GrindEQ__3_} 
\left\{ \begin{array}{c}
\dot{u_1}+i(\frac{\epsilon }{3}+{\epsilon }_1)u_1+{\mathit{\Gamma}}_1u_1-i\left(\frac{3{\gamma }_{11}}{2{\omega }_1}\right)u_1{\left|u_1\right|}^2-i\left(\frac{3{\gamma }_{12}}{{\omega }_1}\right){u^*_1}^2u_2-i\left(\frac{{\widetilde{\gamma }}_{12}}{{\omega }_1}\right)u_1{\left|u_2\right|}^2=0 \\ 
\dot{u_2}+i\epsilon u_2+{\mathit{\Gamma}}_2u_2-i\left(\frac{3{\gamma }_{22}}{2{\omega }_2}\right)u_{2~}{\left|u_2\right|}^2-i\left(\frac{{\gamma }_{21}}{{\omega }_2}\right)u^3_1-i\left(\frac{{\widetilde{\gamma }}_{21}}{{\omega }_2}\right)u_2{\left|u_1\right|}^2={{\tau }_2}/{4}i{\omega }_2 \end{array}
\right. 
\end{equation} 
where $\epsilon = {\omega }_d - {\omega }_2$ and ${\epsilon }_1 = {\omega }_2/3 - {\omega }_1$. Parameters ${\tau}_{2}$, ${\gamma }_{11}$, ${\gamma }_{22}$, ${\gamma }_{12,21}$ and ${\widetilde{\gamma }}_{12,21}$ are determined by ${\tau}_{d}$, $\gamma_{1}$, $\gamma_{2}$, $\gamma$ and ${\widetilde{\gamma }}$ in Eq.~\eqref{GrindEQ__2_} respectively, taking also into account contributions from the renormalization due to the coupling between the two modes. \color{\cl} We note that the rotating wave approximation is justified as the nonlinear terms in the potential energy are much smaller ($\sim\!10^{-7}$) than the harmonic terms. \color{black}

\noindent
\allowbreak

\section{III. ENERGY TRANSFER INTO PERIOD-TRIPLED STATES}

\begin{figure}
\includegraphics[width=3.7in]{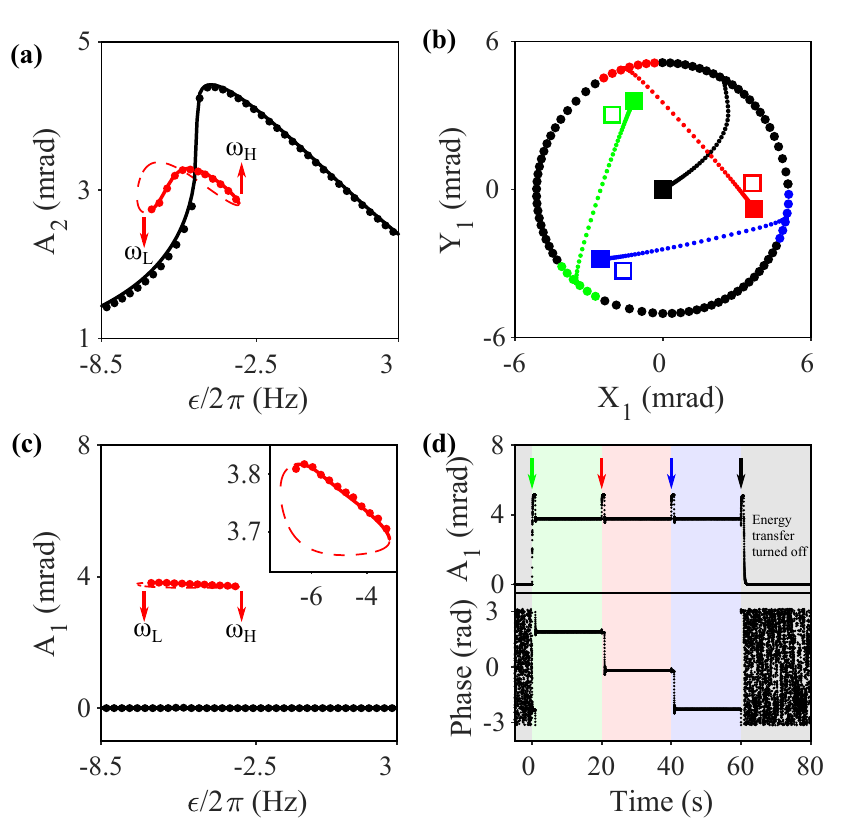}
\caption{\label{Fig02}  (a) Amplitude A${}_{2}$ of mode 2 as a function of the frequency detune $\epsilon ={\omega }_d-{\omega }_2$ of the drive on mode 2 at pump voltage of 6 mV and mode frequency mismatch of  ${\epsilon}_{1}$/2$\pi$ = -16.4 Hz. The solid and dashed lines are stable and unstable vibration states respectively, calculated using Eq.~\eqref{GrindEQ__3_}. Circles are measured results. For the black data, mode 1 is in the zero-amplitude state. For the red data, mode 1 is in the red period-tripled state. The behavior of mode 2 is measured to be identical, within detection noise, regardless of whether mode 1 is in the red, green or blue period-tripled state.  (b) Under the perturbation of a secondary drive applied to mode 1, vibrations of mode 1 are excited, the phase of which depends on the phase of the secondary drive. The initial state lies approximately on a circle in the phase space of mode 1. The color of the large circles indicates to which final state (squares) mode 1 settles after removal of the secondary drive. For each of the four final states, a trajectory is shown (small circles) for one chosen initial state. The hollow squares denote the three saddle points calculated with Eq.~\eqref{GrindEQ__3_}. (c) Amplitude A${}_{1}$ of mode 1 versus the frequency detune $\epsilon$ with the same color scheme as (a). If $\epsilon$ is swept beyond ${\omega}_{L}$ or ${\omega}_{H}$, the response jumps to the black curve. Inset: Zoom in for the isolated loop. (d) Pulses of the secondary drive are applied at time t = 0 s, 20 s and 40 s (arrows) to change the vibration phase of mode 1. Energy transfer from mode 2 is turned off at t = 60 s.}
\end{figure}

 We focus on the steady state oscillations, with ${\dot{u}}_1={\dot{u}}_2=0$. The zero-amplitude state of mode 1 is found to be stable for all ${\tau}_{2}$, $\epsilon$ and ${\epsilon}_{1}$. Measurements on the device yield results that are consistent with this theoretical finding: if the vibration amplitude of mode 1 is initially zero, finite-amplitude vibrations of this mode cannot be excited when ${\omega}_{d}$ is swept up or down. In this scenario, the response of mode 2 is identical to that of a single mode Duffing oscillator, as illustrated by the black data in Fig.~\ref{Fig02}(a) for pump voltage of 6 mV. V${}_{dc1}$ and V${}_{dc2}$ are chosen to be 1.475 V and 1.975 V respectively for the system to be near internal resonance, with ${\epsilon}_{1}/2\pi$ = -16.4 Hz. To achieve stable period-tripled oscillations, it is necessary to apply a proper perturbation to the system to excite mode 1 out of the zero-amplitude state. We choose to perturb the system by applying a secondary ac drive (with amplitude 23 mV) to mode 1 to produce a torque of the form ${\tau }_1\mathrm{cos}\mathrm{}(\frac{{\omega }_d}{3}t+\phi )$. ${\tau}_{1}$ and $\phi$ can be adjusted to excite vibrations in mode 1 with different amplitude and phase, as shown in Fig.~\ref{Fig02}(b) where ${{\theta }_1 \mathrm{=} X_1\mathrm{cos} \left(\frac{{\omega }_d}{3}t\right)+{Y_1\mathrm{sin} \left(\frac{{\omega }_d}{3}t\right)}}$. Upon removal of the secondary drive, the system either settles into one of the three period-tripled states, represented by the red, blue and green squares, or the zero-amplitude state represented by the black square, depending on ${\tau}_{1}$ and $\phi$. Until the perturbation is applied again, mode 1 remains in one of these 4 stable states. 

We emphasize that Fig.~\ref{Fig02}(b) plots the behavior of mode 1 only. In analyzing the dynamics of the system, mode 2 also needs to be included. Hence the phase space is four dimensional (4D) in the rotating frame. To settle into a specific state, the system must start inside the basin of attraction of the corresponding state in 4D phase space. Trajectories from different initial points do not intersect with each other in the 4D space. Figure \ref{Fig02}(b) shows only a projection of the trajectories on the two dimensional space of X${}_{1}$ and Y${}_{1}$.

Figure \ref{Fig02}(c) shows that the tri-stable branch of finite amplitude vibrations of mode 1 (shown in red) is disconnected from the zero-amplitude branch. The corresponding isolated branch of mode 2 is shown in red in Fig.~\ref{Fig02}(a). When $\epsilon$ is increased beyond ${\omega}_{H}$, mode 1 jumps to the zero-amplitude state. Correspondingly, mode 2 jumps to the response of an ordinary Duffing oscillator. The system also undergoes a jump if $\epsilon$ is decreased beyond ${\omega}_{L}$. Once a jump to the zero-amplitude state takes place, getting back to any one of the three period-tripled states in mode 1 requires activation, as discussed earlier. The amplitude for the unstable oscillation states (the saddle points) is calculated from Eq.~\eqref{GrindEQ__3_} and is shown as dashed lines (Supplementary Information). We find that a saddle-node bifurcation occurs at ${\omega}_{H}$, where one stable state merges with one saddle point and disappear. At ${\omega}_{L}$, the system undergoes a Hopf bifurcation where limit cycles develop in the rotating frame, as we will discuss later. In mode 1, the branches of stable and unstable states form an isolated loop. For mode 2, the stable and unstable branches attain peak amplitude at $\epsilon/2\pi$ = -5.5 Hz and -6.8 Hz respectively. If the amplitude of the pump is reduced, the loops decrease  in size and eventually shrink to zero at pump amplitude of 5.5 mV. Below this threshold value of pump amplitude, only the zero-amplitude state is stable and the period-tripled states cannot be excited even with activation.

The energy transfer from mode 2 to mode 1 can be controllably turned on and off. Figure \ref{Fig02}(d) plots the vibration amplitude and phase of mode 1 measured as a function of time. Initially, mode 1 is in the zero-amplitude state with no energy transferred from mode 2. The phase fluctuates due to thermal motion. At time t = 0, the secondary drive with $\phi $ = 150${}^{o}$ is turned on for a duration of 0.9 s and then turned off. With this perturbation, mode 1 settles to the green period-tripled state. At t = 20 s and 40 s, similar secondary drive pulses are applied with $\phi $ = 30${}^{o}$ and 270${}^{o}$ respectively, putting mode 1 in the red period-tripled state followed by the blue one. Energy transfer from mode 2 is turned off when a pulse of the secondary drive with $\phi $ = 90${}^{o}$ is applied at t = 60 s. The vibration of mode 1 goes back to zero and the phase fluctuates randomly.

\noindent

\section{IV. TWO COEXISTING SETS OF PERIOD-TRIPLED STATES}

\begin{figure}
\centering
    \includegraphics[width=3.4in]{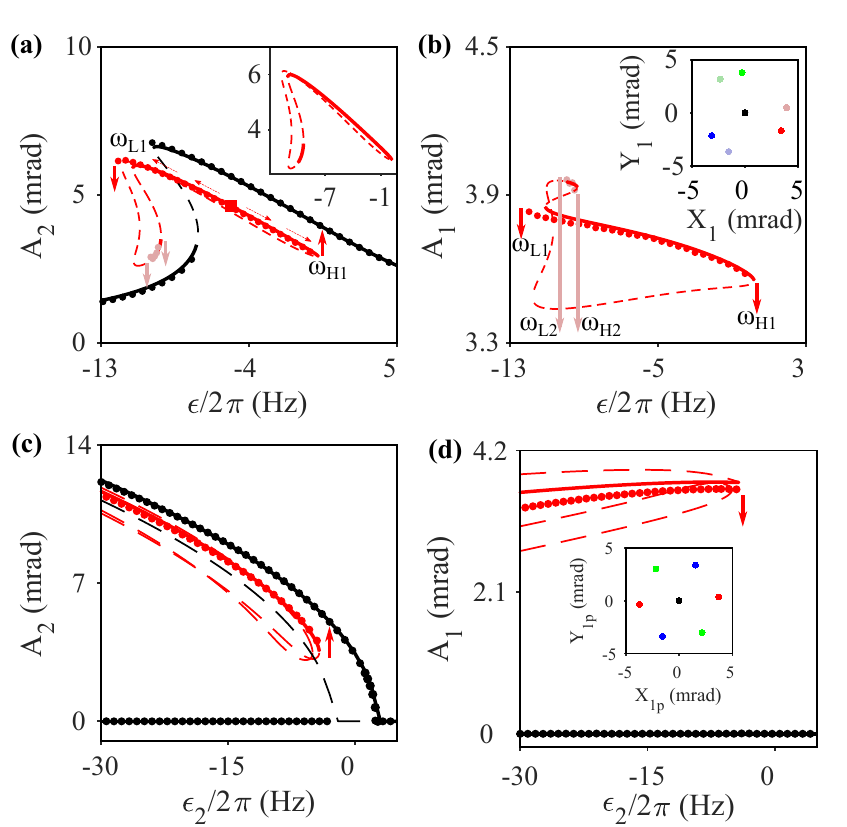}
    \caption{(a) Dependence of the vibration amplitude of mode 2 on the frequency detune of the resonant drive. Black circles are measured amplitude of mode 2 when mode 1 is in the zero-amplitude state. Light and dark red circles represent the measured amplitude of mode 2 when mode 1 is in a period-tripled state. Solid lines are stable states calculated with Eq.~\eqref{GrindEQ__3_}. Dashed lines are unstable states. The left and right light red arrows mark ${\omega}_{L}$${}_{2}$ and ${\omega}_{H2}$${}_{\ }$respectively. Inset: For clarity, calculated amplitudes are shown without measurement data. (b) Vibration amplitude of mode 1. Inset: Two sets of period-tripled states with different amplitudes plotted in the phase space. (c) The vibration amplitude of mode 2 when its eigenfrequency is parametrically modulated near 2${\omega}_{2}$. (d) The corresponding vibration amplitude of mode 1. Inset: Six coexisting states with identical amplitude but with phases differing by $\pi$/3 in phase space.}
    \label{Fig03}
\end{figure}

 As discussed earlier, vibrations in mode 2 results in an effective parametric modulation of mode 1 through the term proportional to 3${\theta}_{1}^2{\theta}_2$ in Eq.~\eqref{GrindEQ__2_}. This approach of generating period-tripled oscillations via internal resonance offer a number of advantages over directly modulating the cubic term in the confinement potential of a single mode [Eq.\eqref{GrindEQ__1_}]. In addition to the resonant enhancement of the modulation amplitude via the response of mode 2, multiple sets of period-tripled states can be generated. Since each stable oscillation state in mode 2 can give rise to one set of stable period-tripled oscillation states in mode 1, the occurrence of multi-stable oscillation states of mode 2 opens the possibility for the coexistence of multiple sets of period-tripled states in mode 1 with different amplitudes and/or phase.

We first consider the case when the resonant drive on mode 2 is increased so that bistability develops in the isolated loop of this mode for a certain range of $\epsilon$ as shown in Fig.~\ref{Fig03}(a) when the pump voltage is increased to 9 mV. The peak in the isolated branch of stable oscillations (red) becomes higher and covers a wider range of frequencies. It bends over towards low frequencies in a manner similar to the ordinary Duffing response of mode 2 (black). As a result, a new, isolated branch of stable vibrations emerges at lower amplitude, plotted in light red in Fig.~\ref{Fig03}(a). The isolated branch of unstable states, calculated using Eq.~\eqref{GrindEQ__3_}, also becomes wider and higher. It is connected to the stable branch as shown in the inset of Fig.~\ref{Fig03}(a). For the upper branch of stable vibrations, there are saddle-node bifurcations at both ends (${\omega}_{L1}$ and ${\omega}_{H1}$). Starting from ${\epsilon}/2{\pi}$ = -5.1 Hz [red square in Fig~\ref{Fig03}(a)], if the drive frequency is swept up beyond ${\omega}_{H1}$, as indicated by the thin dotted arrows pointing to the right, mode 2 jumps up to the Duffing response of a single mode (black). If, instead, the driving frequency is swept down, mode 2 jumps down to the Duffing response at ${\omega}_{L1}$.For the lower branch of stable vibrations, there is a saddle-node bifurcation on the right end at ${\omega}_{H2}$. The left end at ${\omega}_{L2}$ is a Hopf bifurcation where limit cycles develop in the rotating frame, as we will discuss later. Starting from a stable vibration state in the lower branch, sweeping the drive frequency up or down eventually lead to jumping of mode 2 down to the Duffing response, as indicated by the two light red arrows. Once mode 2 jumps to the Duffing response from either the upper or lower branch, merely reversing the direction of frequency sweeping does not bring the system back to the isolated branch. Returning to the isolated branch requires activation, similar to the case of a single set of period-tripled states discussed earlier. 

Each stable state in the isolated loop of mode 2 is associated with one set of three stable period-tripled oscillation states of mode 1. Figure \ref{Fig03}(b) shows the amplitude of the period-tripled states in mode 1. The zero-amplitude state, despite lying out of the range of the vertical axis of the plot, remains stable. For driving frequencies between ${\omega}_{L2}$ and ${\omega}_{H2}$ there are a total of 7 stable states in mode 1. The inset of Fig.~\ref{Fig03}(b) shows the seven coexisting states in phase space for $\epsilon/2\pi$ = -9.72 Hz. We note that no switching between the states is observed. Thermal noise is not strong enough to induce transitions in the duration of our measurement. To place mode 1 in the low-amplitude branch, one method is to start from a smaller drive amplitude on mode 2 that yields only one single branch of period-tripled states, as in Fig.~\ref{Fig02}. When the drive voltage is gradually increased, the system settles to the low-amplitude branch in mode 1. Placing mode 1 in the high-amplitude branch requires slight modifications to the scheme because the frequency range for stable period-tripled states (between ${\omega}_{L2}$ and ${\omega}_{H2}$) does not overlap with that at smaller drive amplitude of Fig. 2 (between ${\omega}_{L}$ and ${\omega}_{H}$). The drive amplitude is increased in small steps. Between successive steps, the drive detune frequency ${\epsilon}$ is decreased by an appropriate amount so that mode 1 settles in the high-amplitude branch at the end of the process. More details of preparing mode 1 in the high-amplitude state is described in Appendix D.

 Coexisting stable states of vibrations of mode 2 can also be excited by parametric modulating its resonant frequency. We modulate the gradient ${d\tau }/{d\theta }$ of the electrostatic torque exerted by the electrode on the small plate at frequency ${\omega}_{p}$ near 2${\omega}_{2}$(see Appendix A). When there is no energy transfer to mode 1, mode 2 follows the response of a single-mode parametric resonator, as shown in black in Fig.~\ref{Fig03}(c). The oscillation states are period-doubled with respect to the modulation. They occurs in pairs that are out of phase with each other. For certain range of parameters, internal resonance enables the energy transfer into mode 1, leading to stable vibrations in mode 1 with period $\frac{12\pi }{{\omega }_p}$. Figure \ref{Fig03}(d) plots the dependence of the amplitude of mode 1 on the modulation frequency detuning ${\epsilon}_{2}$ = ${\omega}_{p}$/2 - ${\omega}_{2}$. We measure six coexisting stable vibration states of mode 1, as shown in the inset of Fig.~\ref{Fig03}(d) where ${{\theta }_1\mathrm{=}X_{1p}\mathrm{cos} \left(\frac{{\omega }_p}{6}t\right)+{Y_{1p}\mathrm{sin} \left(\frac{{\omega }_p}{6}t\right)}}$. The six stable states are measured to have phase offset of $\pi$/3 from one another. They are period-tripled compared to oscillations in mode 2, and period-sextupled compared to the parametric modulation on mode 2. Unlike Fig.~\ref{Fig02}(b) and the inset of Fig.~\ref{Fig03}(b), the inset of Fig.~\ref{Fig03}(d) displays six-fold symmetry. When energy is transferred to mode 1, mode 2 follows the red response in Fig.~\ref{Fig03}(c) that is slightly shifted from the black branch. In calculating the vibrations amplitudes and phase, Equations \eqref{GrindEQ__2_} and \eqref{GrindEQ__3_} need to be modified (See Appendix F). 

\noindent
\allowbreak
\section{V. COEXISTING LIMIT CYCLES}

\begin{figure}[h]
 \centering
    \includegraphics[width=3.4in]{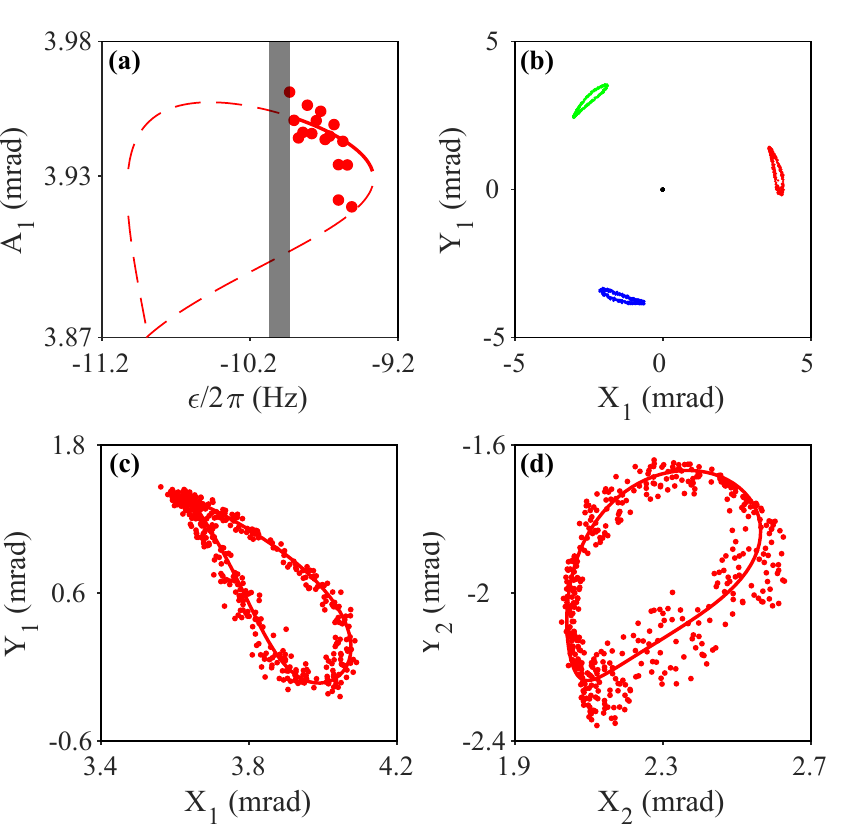}
    \caption{ (a) Zoom-in of Fig.~\ref{Fig03}(b). Stable period-tripled states in mode 1 turn into limit cycles at the right end of the grey band. The limit cycles disappear at the left end of the gray band. (b) Three coexisting limit cycles in mode 1 are measured at frequency detune $\epsilon/2\pi$ of -10.01 Hz. (c) Zoom-in on the red limit cycle. The line represents calculations. (d) Corresponding data for mode 2.}
    \label{Fig04}
\end{figure}

Interesting phenomena such as limit cycles and chaos have been demonstrated in coupled micro- and nanomechanical resonators \cite{PhysRevB.79.165309,PhysRevLett.106.094102,PhysRevLett.109.037205,doi:10.1063/1.5085219,PhysRevLett.125.174301}. In particular, limit cycles are observed for two coupled modes in 3:1 internal resonance when energy is supplied to the lower mode \cite{doi:10.1063/1.5085219}. Our system, in contrast, is under periodic drive on the higher mode and shows qualitatively different behavior. Figure \ref{Fig04}(a) plots the amplitude of mode 1 when mode 2 is driven at pump amplitude of 9 mV. It is a zoom-in of the high-amplitude branch of Fig.~\ref{Fig03}(b). At $\epsilon/2\pi$ = -9.93 Hz, a Hopf bifurcation takes place and limit cycles emerge. Each of the three period-tripled states evolves into a limit cycle in the rotating frame, with a three-fold symmetry illustrated in Fig.~\ref{Fig04}(b). As $\epsilon$ is further decreased, the limit cycles become larger and eventually disappears at $\epsilon/2\pi$ -10.07 Hz. Without energy transfer from mode 2, the vibration amplitude of mode 1 drops to zero. The grey band in Fig.~\ref{Fig04}(a) indicates the frequency range for the occurrence of limit cycles both in measurement and in calculations.  Limit cycles only appear in a narrow frequency range of $\mathrm{\sim}$ 0.14 Hz that is small compared to the frequency range at which stable period-tripled vibrations take place in mode 1 ($\mathrm{\sim}$ 12 Hz). For smaller drive amplitudes in Fig.~\ref{Fig02}(c), limit cycles also develop at the left side of the isolated branch of period-tripled vibrations. However, the frequency range at which limit cycles occur is too small to be visible. 

\section{VI. DISCUSSIONS AND OUTLOOK}

 Our experiment shows that by using an electromechanical resonator with two modes at internal resonance, the nonlinear coupling leads to an effective parametric modulation on the lower mode that is resonantly enhanced by the higher mode. Upon activation, energy can be transferred from the higher mode to the lower mode to excite oscillations in the latter with a period that is tripled the period of pumping. The system can be prepared in any one of the three period-tripled states of mode 1 and remains stable in the duration of the experiment, opening the possibility of using parametrically driven mechanical modes for ternary memory and logic. More efficient schemes for writing bits in period-doubled states in ``parametrons'' \cite{mahboob2008bit,mahboob2011interconnect,nosan2019gate,PhysRevLett.123.254102} can be modified to be used in the period-tripled system. Demonstration of such schemes in our two-mode system is in progress. 

 Away from the internal resonance with the lower mode, the higher mode has multiple stable states with different amplitude and/or phase due to its own nonlinearity. Each of these states can enhance the effective parametric modulation on the lower mode when the two modes are tuned into internal resonance. If the coupling is sufficiently strong, each state will be accompanied by a set of period-tripled states in the lower mode. We demonstrate the excitation of two sets of period-tripled states when the higher mode is under a strong resonant drive. Alternatively, we apply parametric modulation to the higher mode, resulting in two sets of states in mode 1 that are period-tripled with respect to vibrations in mode 2 and period-sextupled with respect to the parametric modulation applied. Mode 2 can also be designed and excited to have even larger number of coexisting stable states. For example, if the Duffing (third order) and fifth order nonlinear coefficients in the restoring force are of opposite signs, the higher mode can possess three stable states at different amplitude and phase \cite{2019Frequency}. If each of these states generates a set of period-tripled states in the lower mode, there will be a total of ten coexisting states in the latter including the zero-amplitude state. A single resonator can therefore, in principle, be used to represent a decimal digit. Alternatively, if mode 2 is subjected to parametric modulation described by Eq.~\eqref{GrindEQ__1_}, three coexisting states can be excited. Internal resonance with the lower mode can potentially lead to nine states in the lower mode with finite vibration amplitude but phase shifted by 2$\pi$/9. With a period that is 9 times the parametric modulation, the reduction of discrete time-translation symmetry will be even more pronounced than the data presented in this paper.

 Near the bifurcation points where stable period-tripled states disappear, effects of fluctuations are expected to be strong. For a single mode system experiencing period-tripling described by Eq.~\eqref{GrindEQ__1_}, the basins of attraction of the period-tripled states are isolated from each other in the 2D phase space. As shown in Fig.~\ref{fig01}(b), each basin is surrounded by the basin of the zero-amplitude state. In the presence of thermal or external noise, the system fluctuates about the stable state. If the system crosses the separatrix, it switches to the zero-amplitude state. For weak noise, the system does not switch to the other two period-tripled states. Interestingly, it has been predicted that the behavior is qualitatively different in quantum oscillators, where direct switching between the period-tripled states can take place \cite{PhysRevE.100.052148}. In the two-mode classical system of this paper, the phase space is 4D, making it difficult to plot and visualize the basin of attraction. By injecting noise into the drive to increase the effective temperature, the system can be induced to switch out of a stable period-tripled state. Preliminary results indicate that when the system switches out of one of the period-tripled states, it settles in the zero-amplitude state for the system parameters chosen in the current experiment. The results are consistent with the notion that the basins of the period-tripled states are isolated from each other in the 4D phase space, similar to the much simpler case of the single mode system described by Eq.~\eqref{GrindEQ__1_}. Whether the behavior changes for other system parameter values, especially those that exhibit limit cycles in phase space, remains an open question that warrants further studies.

The switching from the isolated branch of period-tripled states to zero amplitude in both directions can be used for detecting small perturbations of the eigenfrequencies of the modes. At the saddle-node bifurcations discussed in this paper, the activation barrier for switching is expected to scale with the frequency detuning from the bifurcation point with an exponent of 3/2 \cite{Theory1984,PhysRevLett.94.156403,PhysRevB.73.172302}. Unlike traditional bifurcation amplifiers \cite{PhysRevLett.93.207002} that are only sensitive to unipolar change in parameters, here perturbations in both directions lead to jumps in amplitude that can be easily detected. High sensitivity can be achieved by tuning the amplitude of the pump on the mode 2 to produce a narrow frequency range at which the isolated period-tripled branch in the lower mode is stable.

 With the capability to excite and detect vibrations of the two modes independently, our system provides a versatile platform for studying nontrivial nonlinear effects including limit cycles and dynamical chaos that are absent from single mode systems with only two dynamic variables described by Eq.~\eqref{GrindEQ__1_}. 
We demonstrated that a set of 3 coexisting limit cycles develops over a narrow range of driving frequencies when mode 2 is under resonant drive. \color{\cl} The limit cycles in the rotating frame correspond to non-sinusoidal vibrations that could lead to the development of frequency combs \cite{mahboob2012tuneable,PhysRevLett.118.033903,PhysRevLett.121.244302}. \color{black} Just before the limit cycles disappear, fluctuations are expected to become strong. Future efforts will explore whether period doubling cascades and/or chaotic motion arise when the driving amplitude on mode 2 is further increased. 
 
 \color{\cl} While the period-tripled vibrations reported here are observed in two vibrational modes in a micromechanical resonator designed to be near internal resonance, the findings are generic and applicable to other resonators, such as coupled superconducting coplanar waveguides \cite{PhysRevB.96.174503}, optical cavities \cite{PhysRevLett.80.492}, membranes of two-dimensional materials \cite{samanta2015nonlinear,guttinger2017energy} and carbon nanotubes \cite{eichler2012strong}. The main requirements are that the ratio of the frequencies of the two nonlinearly coupled modes are near 3:1 and the higher mode is subjected to a resonant drive sufficiently strong but well-below the threshold of driving the system into dynamical chaos. In fact, the coexistence of period-tripled oscillation states and the zero-amplitude state was demonstrated in superconducting coplanar waveguide resonators \cite{PhysRevB.96.174503}. However, the detected signal includes contributions from all four states as the system switches among them due to noise. 
 
 The interplay of fluctuations and dynamics has novel, and yet fairly general features in systems that display period tripling. Some of these features arise because the zero-amplitude state remains dynamically stable independent of the drive frequency and strength.  Activation is required to excite the system into one of the three period-tripled states. The activation can come from thermal noise \cite{tadokoro2020noise} or, as recently suggested \cite{PhysRevLett.128.187701}, from quantum fluctuations. A nontrivial aspect of the activation is that, as the driving amplitude is increased, the unstable states in the phase space [hollow squares in Fig. 1(b)] move toward the origin (the zero-amplitude state). The activation barrier for switching out of the zero-amplitude state approaches zero, but never disappears completely. It has been predicted that in this limit the switching dynamics is qualitatively different from the escape from metastable states near bifurcation points: in contrast to conventional wisdom, there is no detailed balance and the switching is not controlled by a soft mode. Even though the theoretical results \cite{tadokoro2020noise,PhysRevLett.128.187701} refer to the case of a single mode, the analysis can likely be extended to the coupled modes. It is possible that such generic behavior in switching could be revealed in resonator systems with modes at internal resonance \cite{eichler2012strong,samanta2015nonlinear,guttinger2017energy,PhysRevB.96.174503,PhysRevLett.80.492}, including the one measured in this paper.
 
Another important feature of period tripling, which makes it qualitatively different from the seemingly similar and extensively studied period doubling, is the nontrivial topology. This topology has been found in the analysis of quantum tunneling between period-three states of a single mode and were related to the geometric phase between the period-three states \cite{PhysRevA.96.052124,Liang_2018,Guo_2020}. We demonstrated in a coupled-mode classical system that when the upper mode is under resonant or parametric drive, up to six symmetry-breaking vibration states in the lower mode can coexist. This finding could motivate studies of topological effects resulting from tunnelling among these states in a quantum resonator. \color{black}

 During the preparation of this manuscript, the authors learned that the decay and energy exchange of two coupled modes in 3:1 internal resonance is studied by Wang \textit{et al}. The main difference from our experiment is that the drive on the higher mode is turned off. Under certain conditions, as the two modes decay, their vibrations are found to phase lock with each other so that the period of the lower mode is tripled that of the upper mode over a certain duration. Period tripling therefore plays an important role in both the steady state and transient response of coupled mechanical modes. It holds promise in providing a new platform for using mechanical resonators for subharmonic generation and sensing applications. 

%	\emph{Acknowledgments.}---	

\section{ACKNOWLEDGEMENTS}

This work is supported by the Research Grants Council of Hong Kong SAR, China (Project No. 16305117). We acknowledge useful discussions with M. I. Dykman.

\begin{appendix}

	\renewcommand{\theequation}{A-\arabic{equation}}
	\setcounter{equation}{0}  

	\section{APPENDIX A: GENERATION OF ELECTROSTATIC TORQUE AND DETECTION OF VIBRATIONS} \label{APPENDIX A}
	For each plate, a voltage $V_d$ applied to the left electrode in Fig.~\ref{fig01}(d) generates an electrostatic torque:  
	\begin{equation} \label{Eq_A1} 
     {\tau}=\frac{1}{2}\frac{dC}{d{\theta}}V_{d}^{2} 
     \end{equation}
    where $C$ is the capacitance between the movable plate and the electrode. The torque can be written as a Taylor expansion about the equilibrium angle $\theta_0$:
    \begin{equation} \label{Eq_A2} 
     {\tau}=\frac{1}{2}[C'(\theta_{0})+C''(\theta_{0})(\theta-\theta_{0})+\frac{1}{2}C^{(3)}(\theta_{0})(\theta-\theta_{0})^{2}+\frac{1}{6}C^{(4)}(\theta_{0})(\theta-\theta_{0})^{3}]V_{d}^{2} 
     \end{equation}
     where $C'$, $C''$, $C^{(3)}$ and $C^{(4)}$ represent the first, second, third and fourth derivative of C with respect to $\theta$ respectively. Higher order terms are neglected. The voltage $V_{d}$ consists of a sum of a dc component $V_{dc}$ and an ac component with amplitude $V_{ac}$ and frequency $\omega_{f}$:
    \begin{equation} \label{Eq_A3} 
     V_{d}=V_{dc}+V_{ac}cos(\omega_{f}t)
     \end{equation}where $V_{ac} << V_{dc}$.
 
    The time-independent component of $\tau$ given by Eqs.~(\ref{Eq_A2}) and (\ref{Eq_A3}) changes the system parameters. First, the constant term produces a shift in the equilibrium position of $\theta$ to $\theta_{0}$. Second, electrostatic spring softening is induced by the term proportional to $\theta-\theta_{0}$. The restoring torque on the plate is modified by an amount $\frac{1}{2}C''(\theta_0)V_{dc}^{2}(\theta-\theta_0)$. Our experiment relies on such electrostatic tuning of the resonant frequencies to control the frequency mismatch $\epsilon_{1}$ of the two modes by adjusting $V_{dc1}$ and $V_{dc2}$. Third, contributions from the nonlinear restoring torques of $\frac{1}{4}C^{(3)}(\theta_{0})V_{dc}^{2}(\theta-\theta_{0})^{2}$ and $\frac{1}{12}C^{(4)}(\theta_{0})V_{dc}^{2}(\theta-\theta_{0})^{3}$ dominate the Duffing nonlinearity.
    
     The ac voltage leads to both an additive torque $C'(\theta_{0})V_{dc}V_{ac}cos(\omega_{f}t)$ and a torque $C''(\theta_{0})V_{dc}V_{ac}cos(\omega_{f}t)(\theta-\theta_{0})$ that modulates the effective spring constant. In the main text, when the drive is applied to mode 2 at a frequency is close to $\omega_{2}$, $\omega_{f}$ is replaced by $\omega_d$ to denote resonant driving. When the frequency is close to $2\omega_{2}$, the main effect is parametric modulation of the spring constant. $\omega_{f}$ is replaced by $\omega_{p}$ to distinguish the parametric modulation from the case of resonant driving.
     
     Vibrations of each plate are detected by capacitive detection. The application of dc voltages to the right electrodes in Fig.~\ref{fig01}(d) leads to build-up of charges between the top plates and the electrodes. Rotations of the plates lead to changes in their capacitance. Charge flowing out of the two plates are measured simultaneously and independently by two separate charge sensitive amplifiers, the outputs of which are fed into the two channels of a lock-in amplifier (Zurich Instruments HF2LI).
     
     \renewcommand{\theequation}{B-\arabic{equation}}
    \setcounter{equation}{0}  

	\section{APPENDIX B: CALCULATION OF STABLE VIBRATION STATES AND SADDLE POINTS} \label{APPENDIX B}

	The stable states of vibration and the saddle points are calculated by assuming that the complex amplitudes ${u}_{1}$ and ${u}_{2}$ are independent of time. Equations \eqref{GrindEQ__3_} reduce to two algebraic equations, the solutions of which give the stationary states:

\begin{equation} \label{Eq_B1} 
\left\{ \begin{array}{c}
(\frac{\epsilon }{3}+{\epsilon }_1)|u_1|-i{\mathit{\Gamma}}_1|u_1|-\left(\frac{3{\gamma }_{11}}{2{\omega }_1}\right){\left|u_1\right|}^3-\left(\frac{{\widetilde{\gamma }}_{12}}{{\omega }_1}\right)|u_1|{\left|u_2\right|}^2=\left(\frac{3{\gamma }_{12}}{{\omega }_1}\right){|u_1|}^2|u_2|e^{i(\vartheta_2-3\vartheta_1)} \\ 
\epsilon|u_2|-i{\mathit{\Gamma}}_|2u_2|-\left(\frac{3{\gamma }_{22}}{2{\omega }_2}\right){\left|u_2\right|}^3-\left(\frac{{\gamma }_{21}}{{\omega }_2}\right)|u|^3_1{e}^{i(3\vartheta_1-\vartheta_2)}-\left(\frac{{\widetilde{\gamma }}_{21}}{{\omega }_2}\right)|u_2|{\left|u_1\right|}^2=-\left(\frac{\tau_2}{4\omega _2}\right){e}^{-i\vartheta_2} \end{array}
\right. 
\end{equation} 
where ${u}_{1,2}$ = $|{u}_{1,2}|{e}^{i\vartheta_{1,2}}$. The stability of the states is found by linearizing Eq.~\eqref{GrindEQ__3_} about the stationary values of ${u}_{1}$ and ${u}_{2}$.

\renewcommand{\theequation}{C-\arabic{equation}}
\setcounter{equation}{0}  
\section{APPENDIX C: DETERMINATION OF SYSTEM PARAMETERS} \label{APPENDIX C}

      \begin{figure}
   
     \includegraphics[width=3.4in]{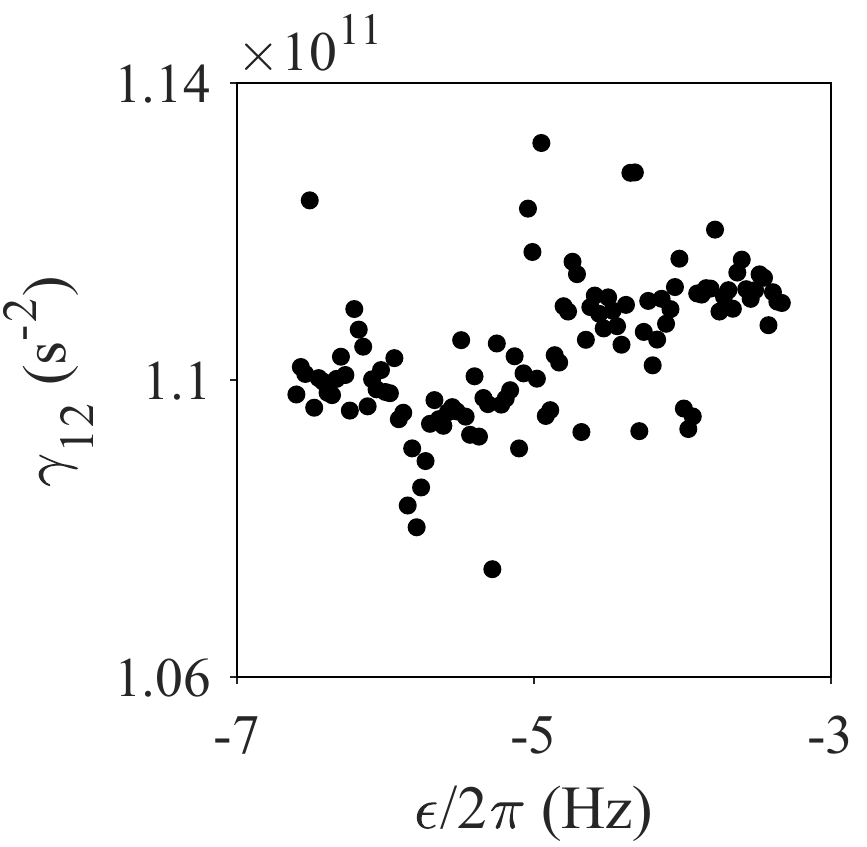}
      \caption{\label{Fig05} Dependence of $\gamma_{12}$ on the drive detune $\epsilon$ calculated with Eq.~(\ref{Eq_C2}). The measured values of $u_{1,2}$ from the isolated branches of Figs.~\ref{Fig02}(a) and \ref{Fig02}(c) are used.}
      \end{figure}
   
   All parameters in Eq.~(\ref{GrindEQ__3_}) are individually characterized without using the data for period-tripled vibrations, with the exception of $\gamma_{12,21}$ that originate from the coupling energy $\gamma q_{1}^{3}q_{2}$. $\gamma_{12,21}$ are fitted using the vibration amplitudes of the two modes in Figs.~\ref{Fig02}(a) and \ref{Fig02}(c) when period-tripled vibrations take place in mode 1. There are no fitting parameters for the theory curves in Figs.~\ref{Fig03} and \ref{Fig04}.
   
   The eigenfrequency $\omega_{1}$ and damping coefficient $\Gamma_{1}$ of mode 1 are determined by fitting to the response when a small resonant drive is applied to mode 1 only but not to mode 2. Subsequently, the driving amplitude is increased so that the response becomes nonlinear. The constant $\gamma_{11}$ ($-1.51\times10^{12}$ $s^{-2}$) in the $u_{1}|u_{1}|^{2}$ Duffing nonlinear term is obtained from the shift of the peak frequency as the vibration amplitude increases. The procedure is then repeated for mode 2 to determine $\omega_{2}$, $\Gamma_{2}$ and $\gamma_{22}$ ($-8.04\times10^11$  $s^{-2}$).
   
    Dispersive coupling energy of the form  $\frac{1}{2}\gamma\theta_{1}^{2}\theta_{2}^{2}$ leads to a shift in the resonant frequency of one mode by an amount proportional to the square vibration amplitude of the other mode. In Figs.~\ref{fig01}(e) and \ref{fig01}(f), the slopes of the linear fits are given by $\frac{\widetilde{\gamma}_{12}}{4\omega_{1}}$ and $\frac{\widetilde{\gamma}_{21}}{4\omega_{2}}$  respectively, yielding  ${\widetilde{\gamma}_{12}} = -1.92 \times 10^{11}$ $s^{-2}$ and ${\widetilde{\gamma}_{21}} \approx -1.34 \times 10^{12}$ $s^{-2}$. Measurement of ${\widetilde{\gamma}_{12}}$ and ${\widetilde{\gamma}_{21}}$ also allows the ratio of the effective moment of inertia of the two modes to be determined, through the relation $\frac{I_2}{I_1} = \frac{{\widetilde{\gamma}_{12}}}{{\widetilde{\gamma}_{21}}}$. In mode 2, the rotation of the large plate is negligible. By taking $I_2$ to be the moment of inertia of the small plate about the rotation axis ($7.61\times10^{-20}\ kg\ m^2$), $I_1$ is calculated to be $5.34\times10^{-19}\ kg\ m^2$.
    
   All the system parameters in Eq.~(\ref{GrindEQ__3_}) are thus individually characterized. The only exceptions are $\gamma_{12}$ and $\gamma_{21}$. Both originate from the coupling energy $\gamma q_1^3q_2$. Their ratio is known, given by $\frac{\gamma_{12}}{\gamma_{21}}=\frac{I_2}{I_1} $, similar to the case for dispersion coupling discussed above. By setting $\dot{u_1}=0$ in the first equation in Eq.~(\ref{GrindEQ__3_}), the steady state vibration amplitudes satisfy:
   \begin{equation} \label{Eq_C1} 
     (\Gamma_1^2+(\frac{1}{3}\epsilon+\epsilon_1-\frac{3\gamma_{11}}{2\omega_1}|u_1|^2-\frac{\widetilde{\gamma}_{12}}{\omega_1}|u_2|^2)^2)=(\frac{3\gamma_{12}}{\omega_1})^2|u_1|^2|u_2|^2
     \end{equation}
$\gamma_{12}$ can be written in terms of the other system parameters and $u_{1,2}$:
       \begin{equation} \label{Eq_C2} 
    \gamma_{12}=\frac{\omega_1\sqrt{ (\Gamma_1^2+(\frac{1}{3}\epsilon+\epsilon_1-\frac{3\gamma_{11}}{2\omega_1}|u_1|^2-\frac{\widetilde{\gamma}_{12}}{\omega_1}|u_2|^2)^2)}}{3|u_1||u_2|}
     \end{equation}

  Figure \ref{Fig05} plots $\gamma_{12}$ calculated with Eq. (\ref{Eq_C2}), using the measured values of ${u_{1,2}}$ [red data points in Figs.~\ref{Fig02}(a) and ~\ref{Fig02}(c)] as the drive detune $\epsilon$ is varied. \color{\cl} Detection noise of $u_{1,2}$ leads to small fluctuations of $\gamma_{12}$ about the mean value \color{black} of $1.10\times10^{11}\ s^{-2}$. $\gamma_{21}=\frac{I_1}{I_2}\gamma_{12}$ is determined to be $7.74\times10^{11}\ s^{-2}$. These values of $\gamma_{12,21}$ are used to generate the theory curves in Figs.~\ref{Fig02}, \ref{Fig03} and \ref{Fig04}. Only the data in Fig.~\ref{Fig02} are used in the fitting to determine $\gamma_{21}$. There are no fitting parameters for the theory curves in Figs.~\ref{Fig03} and \ref{Fig04}.
	
\renewcommand{\theequation}{D-\arabic{equation}}
\setcounter{equation}{0}  
\section{APPENDIX D: DEPENDENCE OF PERIOD-TRIPLED VIBRATIONS ON THE DRIVING AMPLITUDE} \label{APPENDIX D}
 
     \begin{figure}[h]

     \includegraphics[width=6in]{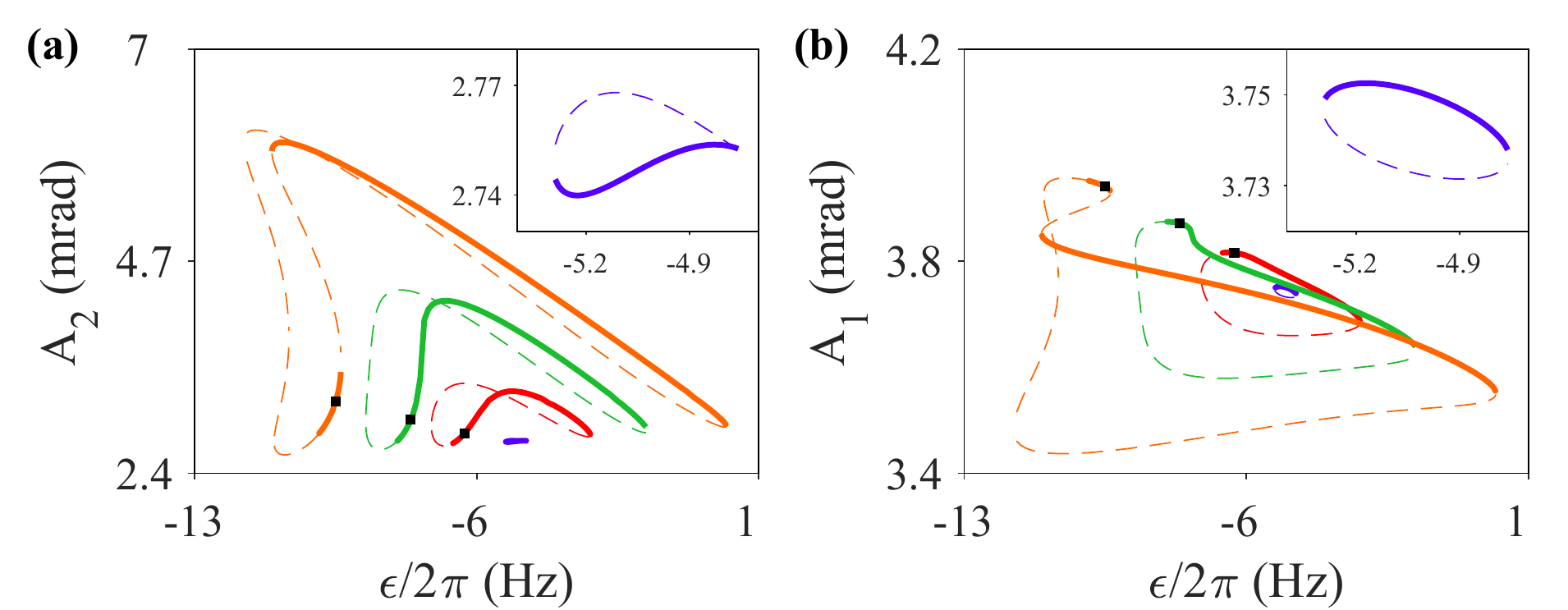}
      \caption{(a) Amplitude $A_2$ of mode 2 as a function of the frequency detune $\epsilon$ of the drive on mode 2 for pump voltages of 5.54 mV (blue), 6 mV (red), 7 mV (green) and 10 mV (orange). Stable and unstable solutions to Eq.~(\ref{GrindEQ__3_}) with $\dot{u_{1}}=\dot{u_{2}}=0$ are plotted as solid and dashed lines respectively. The black squares illustrate one way to access the lower branch for pump voltage of 10 mV: $\epsilon$ is adjusted lower as the drive amplitude is increased. (b) Corresponding vibration amplitude for the period-tripled vibrations in mode 1. Insets: zoom in for pump voltage of 5.54 mV. Solutions for zero vibration amplitude for mode 1 are omitted in both panels.}
      \label{Fig06}
      \end{figure}
      Figure~\ref{Fig06}(a) shows the calculated amplitudes of the stationary points of the isolated loop in mode 2 when the mode is under resonant drive. The calculation is performed by setting $\dot{u_{1}}=\dot{u_{2}}=0$ in Eq.~(\ref{GrindEQ__3_}). Solid and dashed lines represent stable and unstable solutions respectively. As the drive amplitude is increased, the frequency range for stable vibrations widens. The peaks become higher and tilt towards low frequencies. At the largest driving amplitude of 10 mV, the peak tips over and a second, isolated stable branch appears at low amplitudes. Figure~\ref{Fig06}(b) plots the amplitudes for the corresponding vibrations of mode 1 with period tripled that of mode 2. 
      
      Starting from the red curve for a driving amplitude of 6 mV in Fig.~\ref{Fig06}(a), increasing the drive amplitude while keeping frequency detune $\epsilon$ fixed brings mode 2 to the upper branch for driving amplitude of 10 mV. To place mode 2 in the lower branch, it is necessary to decrease the driving frequency as the driving amplitude is increased in small steps. One choice for the driving frequencies is illustrated by the black squares in Fig.~\ref{Fig06}(a).
      
 \renewcommand{\theequation}{E-\arabic{equation}}
\setcounter{equation}{0}  
\section{APPENDIX E: SYSTEM PARAMETERS FOR EXCITNG STABLE PERIOD-TRIPLED STATES} \label{APPENDIX E}
     \begin{figure}[h]

     \includegraphics[width=3.4in]{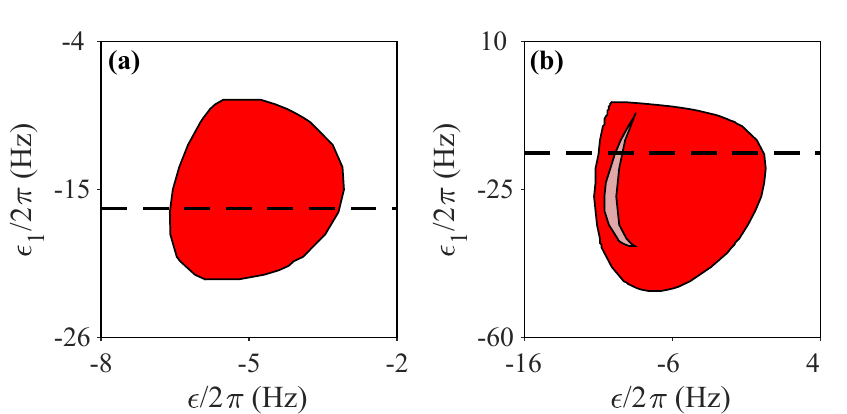}
      \caption{(a) The parameter space of the drive frequency detune $\epsilon$ and the frequency mismatch $\epsilon_1$ of the two modes for pump amplitude of 6 mV. In the red regions, period-tripled oscillations can be excited in mode 1. The zero-amplitude state and the three period-tripled states are stable. In the white regions, only the zero-amplitude state is stable. (b) The pump amplitude is increased to 9 mV. In the light red region, there are two sets of stable period-tripled states in mode 1, together with the stable zero-amplitude state. The horizontal black dashed lines mark $\epsilon_1=-16.4\ Hz$ for the measurements in Figs.~\ref{Fig02} and \ref{Fig03}.}
      \label{Fig07}
      \end{figure}
    The frequency mismatch $\epsilon_1=\omega_2/3-\omega_1$ is fixed for all results presented in the main text, with $\epsilon_1/2\pi=-16.4\ Hz$. We chose $\epsilon_1$ to be non-zero because excitation of period-tripled vibrations in our system requires both modes to vibrate at moderately large amplitudes so that the shifts in eigenfrequencies due to the Duffing nonlinearities are significant. While the Duffing coefficients of both modes are negative, the shift in the frequency of mode 1 is much larger ($\sim$ 5 times) than mode 2. It is therefore necessary to pick $\omega_1$ exceeding $\omega_2/3$ so that at the larger amplitudes, the ratio of the shifted eigenfrequencies becomes close to 3 to satisfy the conditions of internal resonance. For $\epsilon_1>0$, we do not observe any stable period-tripled vibrations in mode 1 in our system.
    
	The range of $\epsilon_1$ that yields stable period-tripled vibrations can be calculated using the procedure described in Appendix B. Figure~\ref{Fig07}(a) shows the range of $\epsilon_1$ and drive frequency detune $\epsilon$ in which period-tripled vibrations of mode 1 are stable, at pump amplitude of 6 mV. When the pump amplitude is increased to 9 mV, it is possible to excite a second branch of period-tripled states, as illustrated by the light red region in Fig.\ref{Fig07}(b).

\renewcommand{\theequation}{F-\arabic{equation}}
\setcounter{equation}{0}  
\section{APPENDIX F: ENERGY TRANSFER TO MODE 1 WHEN MODE 2 IS UNDER PARAMETRIC MODULATION} \label{APPENDIX F}

\begin{figure}[!h]
    \centering
    \includegraphics[width=6in]{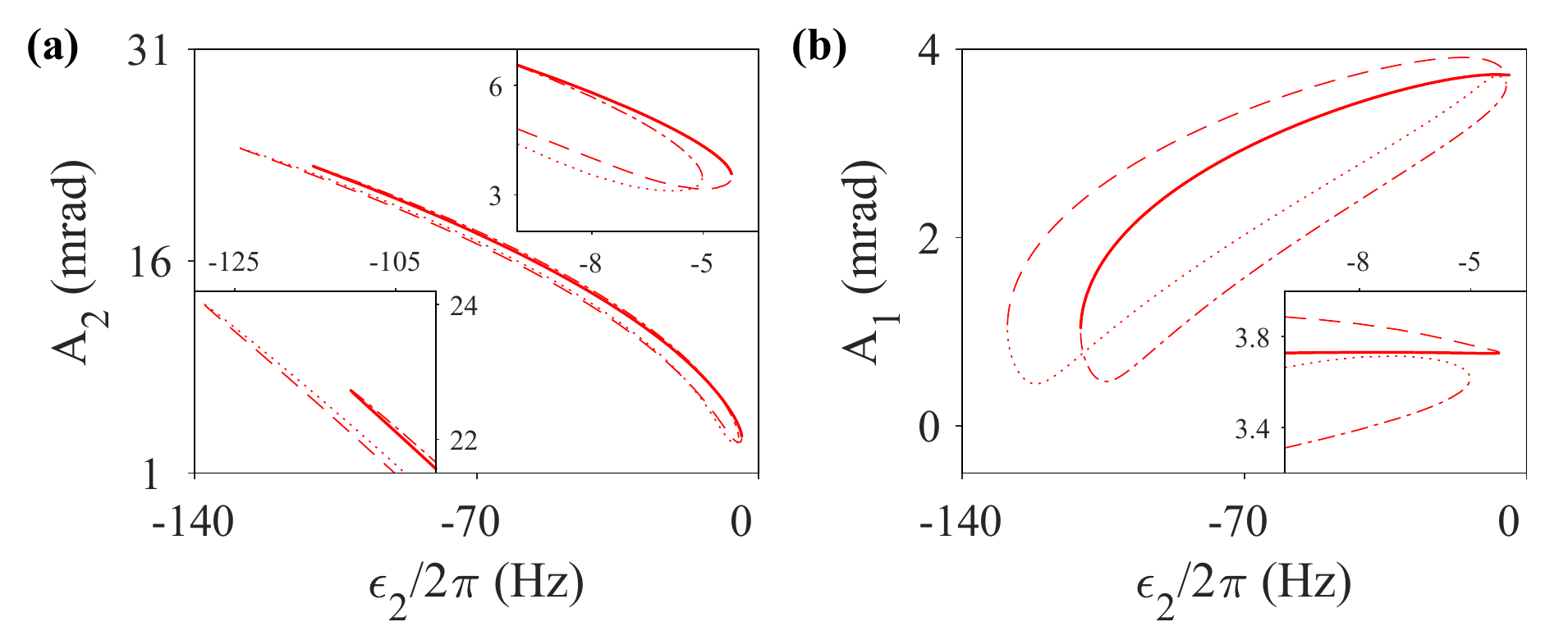}
    \caption{The dependence of the amplitude of the stable vibrations states (solid line) and saddle points (dashed, dotted and dash-dotted lines) of (a) mode 2 and (b) mode 1 on the detuning frequency of the parametric modulation of mode 2. When there is no energy transfer from mode 2 to mode 1, mode 2 displays ordinary parametric resonance and mode 1 remains at zero amplitude. These stationary states are omitted from the plot for clarity. The insets zoom in to show the merging of the branches at the two ends.}
    \label{Fig08}
\end{figure}

When the resonant drive on mode 2 in Eq.~\eqref{GrindEQ__2_} is replaced by parametric modulation of its eigenfrequency, the equation of motions are modified to:
\begin{equation} \label{Eq_B1} 
\left\{ \begin{array}{c}
\ddot{{\theta }_1}+2{\mathit{\Gamma}}_1\dot{{\theta }_1}+{\omega }^2_1{\theta }_1+\frac{{\gamma }_1}{I_1}{\theta }^3_1+3\frac{\gamma }{I_1}{\theta }^2_1{\theta }_2+\frac{\widetilde{\gamma }}{I_1}{\theta }_1{\theta }^2_2=0 \\
\ddot{{\theta }_2}+2{\mathit{\Gamma}}_2\dot{{\theta }_2}+[{\omega }^2_2+\frac{{K}_e}{I_2}{cos \left({\omega }_pt\right)\ }]{\theta }_2+\frac{{\gamma }_2}{I_2}{\theta }^3_2+\frac{\gamma }{I_2}{\theta }^3_1
+\frac{\widetilde{\gamma }}{I_2}{\theta }^2_1{\theta }_2=0 \end{array}
\right. 
\end{equation} 
where $k_{e}=-C^{''} V_{dc2} V_{ac2}$ is the amplitude of the modulation of the torsional spring constant, $C^{''}$ is the second derivative of the capacitance between plate 2 and the driving electrode with respect to ${\theta}_{2}$ and $V_{ac2}$ (80 mV) is the amplitude of the periodic ac voltage applied to the driving electrode. With the modulation frequency ${\omega}_{p}$ close to 2${\omega}_{2}$, we change from ${\theta}(t)$ and $\dot{\theta}(t)$ to complex amplitudes 
${\theta }_1\left(t\right)=u_1\left(t\right)\mathrm{exp}\ \left[i\left({{\omega }_p}/{6}\right)t\right]+\mathrm{c.c.}$.,  ${\theta }_2\left(t\right)=u_2\left(t\right)\mathrm{exp}\ \left[i\left({{\omega }_p}/{2}\right)t\right]+\mathrm{c.c.}$. 
Under the rotating wave approximation, ${u}_{1,2}(t)$ evolves according to:
\begin{equation}\label{Eq_B2} 
\left\{\begin{array}{c}
\dot{u_1}+i(\frac{\epsilon }{3}+{\epsilon }_1)u_1+{\mathit{\Gamma}}_1u_1-i\left(\frac{3{\gamma }_{11}}{2{\omega }_1}\right)u_1{\left|u_1\right|}^2-i\left(\frac{3{\gamma }_{12}}{{\omega }_1}\right){u^*_1}^2u_2-i\left(\frac{{\widetilde{\gamma }}_{12}}{{\omega }_1}\right)u_1{\left|u_2\right|}^2=0 \\
\dot{u_2}+i\epsilon u_2-i\frac{\omega_2k_e}{4I_2}u^*_2+{\mathit{\Gamma}}_2u_2-i\left(\frac{3{\gamma }_{22}}{2{\omega }_2}\right)u_{2~}{\left|u_2\right|}^2-i\left(\frac{{\gamma }_{21}}{{\omega }_2}\right)u^3_1-i\left(\frac{{\widetilde{\gamma }}_{21}}{{\omega }_2}\right)u_2{\left|u_1\right|}^2=0 \end{array}
\right. 
\end{equation}
The stationary states and their stability is determined using the procedure described in Appendix B.

Figure \ref{Fig08}(a) plots stable vibration states and saddle points for mode 2 as solid and dashed lines respectively. For each branch, there are two vibrational states that are opposite in phase and period-doubled with respect to the parametric modulation. The ordinary parametric resonance of mode 2 [black results in Fig.~\ref{Fig03}(a)] is omitted for clarity. Figure \ref{Fig08}(b) shows similar plots for mode 1, where the vibrational states are period-sextupled with respect to the parametric modulation. At both the low and high frequency ends of the stable branch, the stable states merge with the corresponding saddle points in saddle-node bifurcations. 
	
\end{appendix}

\foreach \x in {1,...,5}
{%
\newpage
\includegraphics[scale=0.88,page={\x}]{yan_supplementary.pdf}{\centering}
}

\end{document}